\documentclass[12pt]{article}

\newcommand{\rf}[1]{(\ref{#1})}

\newcommand{\ba}{\begin{array}}
\newcommand{\ea}{\end{array}}
\newcommand{\be}{\begin{equation}}
\newcommand{\ee}{\end{equation}}
\newcommand{\no}{\noindent}
\newcommand{\ods}{\par \vspace{0.5cm} \par}

\newcommand{\const}{{\rm const}}

\newcommand{\e}{{\bf e}}

\newcommand{\R}{{\bf R}}

\newcommand{\scal}[2]{\mbox{$\langle #1 \! \mid #2 \rangle $}}

\newcommand{\MyBox}{\begin{picture}(7,7)(0,0) 
 \put(0,0){\line(1,0){7}} \put(7,0){\line(0,1){7}} 
 \put(0,0){\line(0,1){7}} \put(0,7){\line(1,0){7}} 
 \end{picture} }
\newtheorem{prop}{Proposition}

\newenvironment{Proof}{\par \vspace{2ex} \par
\noindent \small {\it Proof:}}{\hfill \MyBox 
\vspace{2ex} \par } 

\newfont{\cyr}{cmcyr12}

\begin{document}

\title{\bf Discretization of multidimensional 
submanifolds associated with Spin-valued spectral problems}

\author{{\bf Jan L.\ Cie\'sli\'nski}\thanks{Supported by
Polish Committee for Scientific Research: grant KBN 2 P03B 126 22.}
\\ { \footnotesize Uniwersytet w Bia\l ymstoku,
Instytut Fizyki Teoretycznej,}
\\ {\footnotesize 15-424 Bia\l ystok, ul.\ Lipowa 41, Poland }
\\ {\footnotesize e-mail: \tt janek\,@\,alpha.uwb.edu.pl  }
}

\maketitle

\begin{abstract}
\small 
We present a large family of ${\rm Spin}(p,q)$-valued 
 discrete spectral problems. The associated discrete
nets generated by  the so called Sym-Tafel formula are circular nets
(i.e., all elementary quadrilaterals are inscribed into circles).
These nets are discrete analogues of smooth  multidimensional 
immersions in $\R^m$ including isothermic surfaces, Guichard nets,
and some other families of orthogonal nets. 
\end{abstract}

\ods

{\small
 }

\pagebreak

One of the most important 
topics in the classical differential geometry was to study special 
coordinates (nets) on surfaces (and submanifolds) and 
various transformations, associated with the names of Bianchi, B\"acklund, Darboux, 
Ribaucour, Levy, Combescure, Jonas and others \cite{Eis1,Eis2}.
Recently, one can observe a rapid development of a discrete analogue of 
differential geometry of submanifolds, focused on discrete nets and their 
transformations  \cite{BP-Ox,Dol-Chicago,DSM}. 
 Some results in this direction were obtained earlier \cite{Sau}. 
For instance, the discretization of pseudospherical surfaces is known 
since more than 50 years \cite{Wun}.
Now it is clear that the transformations of the classical 
differential geometry (and their discrete analogues) are associated 
with integrable systems of nonlinear partial differential (and difference) 
equations and their soliton solutions
(see \cite{RS,ZMNP}).

In this paper we consider discrete nets, i.e., maps 
$F: {\mathbf Z}^n \rightarrow \R^m$. In the case
$n=2$ they are also called discrete surfaces. 
 The map $\R^n \rightarrow \R^m$, obtained in the continuum 
limit from a discrete net, corresponds to a specific choice of
coordinates on some smooth surface. 

Some examples are in order.
The discrete analogue for asymptotic nets is characterized by the property that any point
$F$ and its all four neighbours ($T_1 F$, $T_2 F$, $T_1^{-1} F$, $T_2^{-1} F$) are co-planar.  
$T_j$ denotes the shift in $j$-th variable, i.e.,
\[
T_j f (m^1,\ldots,m^j,\ldots,m^n) = f (m^1,\ldots,m^j + 1,\ldots,m^n) \ .
\]
Discrete pseudospherical surfaces are defined as discrete asymptotic nets such 
that all segments joining the neighbouring points have equal lengths \cite{BP-pseudo,Wun}.

By an elementary quadrilateral we mean four  neighbouring points: $F$,
$T_k F$, $T_j F$ and $T_kT_j F$.
Planar quadrilaterals correspond in the smooth case  to conjugate nets (i.e., coordinates such that the second fundamental 
form is diagonal) \cite{DS-nets}. 

{\bf Circular nets} (such that every quadrilateral is inscribed into a circle) correspond to curvature lines (i.e., coordinates such 
that both fundamental forms are diagonal) \cite{Bob-kent,CDS}. 

Isothermic immersions (characterized, in the smooth case, by the property that curvature lines admit conformal 
(isothermic) parameterization) in the discrete case are defined by the requirement that the cross-ratio for any elemetary quadrilateral is a negative constant  \cite{BP-izot}.

In this paper, following the procedure applied earlier in the smooth case
\cite{Ci-Cliff}, we identify the 
space $\R^m$ with the vector space $V$  generating 
the Clifford algebra $Cl (V)$: 
\[
      \R^m \ni (x^1, \ldots, x^m) \quad \longleftrightarrow  \quad
       x^1 \e_1 + \ldots + x^m \e_m  \ \in V \subset Cl (V) \ .
\]
We recall that the vector space $V$ equipped with a quadratic form of the signature 
$(p,q)$, $p+q=m$, generates the Clifford algebra $Cl (V) \simeq Cl_{p,q}$. The multiplication in the Clifford algebra $Cl_{p,q}$ satisfies
\cite{Cliff,Lou}
\[ 
 \e_1^2 = \ldots = \e_p^2 = 1  \ , \quad 
 \e_{p+1}^2 = \ldots =  \e_{p+q}^2 = - 1  \ , \quad 
\e_j \e_k = -  \e_k \e_j  \quad (k\neq j) \ ,
 \]
i.e., for any Clifford vectors $v = v^1 \e_1 + \ldots v^m \e_m$ and
$w = w^1 \e_1 + \ldots + w^m \e_m$ we have 
\[
       v w + w v = 2 \scal{v}{w} \equiv  2 (v^1 w^ 1 + \ldots v^p w^p - v^{p+1} w^{p+1} 
 - \ldots - v^{p+q} w^{p+q} ) \ ,
\]
where the right hand side is understood to be proportional to the unit element
${\bf 1}$ of the algebra $Cl (V)$ (in general, we identify scalars, the one-dimensional linear 
space spanned by ${\bf 1}$, with $\R$). 
In particular, the Clifford square of any vector is real.

The algebra $Cl (V)$ is spanned by ${\bf 1}$, vectors $\e_k$ and multi-vectors 
$\e_{k_1}\ldots \e_{k_r}$ ($1 \leq k_1 < k_2 < \ldots < k_r \leq p+q$, $1 < r \leq p+q$) and 
$\dim Cl_{p,q} = 2^{p+q}$.

The Lipschitz group $\Gamma (V)$ (known also as the Clifford group) is 
the multiplicative group (with respect to the Clifford product) 
 generated by Clifford vectors. The group generated by
unit vectors is called ${\rm Pin} (V)$, the group generated by even number of vectors
is denoted by $\Gamma_0 (V)$, and, finally, the group generated by
even number of unit vectors is called ${\rm Spin} (V)$ \cite{Cliff,Lou}. Obviously,
\[
    {\rm Spin} (V) \subset {\rm Pin} (V) \subset \Gamma (V) \subset Cl (V) \ ,
\quad  V \subset {\rm Pin} (V) \ , \quad \Gamma_0 (V) \subset \Gamma (V) \ .
\]

A convenient way to describe circular nets is the {\bf cross ratio} (see, for
example, \cite{Ah}), and especially its generalization for Euclidean
spaces (``the Clifford cross ratio'') \cite{Ci-cross}. Namely, for any sequence 
of 4 points in a Euclidean space we define
\be  \label{cross}
Q(X_1,X_2,X_3,X_4) := (X_1-X_2) (X_2-X_3)^{-1} (X_3-X_4) (X_4-X_1)^{-1}  .
\ee
In general $Q(X_1,X_2,X_3,X_4)$ is an element of $\Gamma_0  (V)$.
In the pseudo-Euclidean case ($pq \neq 0$) 
 there exist non-invertible (isotropic) 
vectors and, therefore, the cross-ratio is not always well defined.
 One can easily 
show the following proposition (\cite{Ci-cross}, compare \cite{BP-izot}).

\begin{prop} The Clifford cross ratio 
$Q(X_1,X_2,X_3,X_4)$ is real (i.e., proportional to the unit element of $Cl (V)$) if and only if $X_1,X_2,X_3,X_4$ lie on a circle or
are co-linear.
\end{prop}

Therefore the Clifford cross-ratio  can be used to characterize 
discrete analogues of curvature nets, isothermic surfaces etc.
The sides of the elementary quadrilateral are given by $D_k F$, $D_j F$, $T_k D_j F$, $T_j D_k F$,
where $D_k F := T_k F - F$. We define
\be
  Q_{kj} (F) := Q (F, T_k F, T_{kj} F, T_j F) = (D_k F) (T_k D_j F)^{-1} (T_j D_k F) (D_j F)^{-1} \ ,
\ee
and formulate the following corollary.  

\begin{prop} 
The net $F=F(m^1,\ldots,m^n)$ is a circular  net if and only if $Q_{kj} (F) \in \R$
for any $k,j \in
\{1,\ldots,n\}$.
\end{prop}

An interesting connection between submanifolds (or discrete nets) and integrable systems
is provided by the Sym-Tafel formula $ F = \Psi^{-1} \Psi,_{\lambda}$ 
\cite{Ci-FG,S-1,Sym}, where $\Psi$ is a solution of some linear 
problem (``Lax pair'') with the spectral parameter $\lambda$. 
This formula was applied in order to discretize  
pseudospherical and isothermic surfaces \cite{BP-izot, BP-pseudo}.

\begin{prop}   \label{long}
We consider the Clifford algebra $Cl (V\oplus W)$, where $V, W$ are  vector 
spaces ($\dim V = q, \dim W = r$) equipped with quadratic forms. 
Let $\Psi$ is a solution of the following discrete linear 
problem:
\be  \label{as1}
T_j \Psi = U_j \Psi \ ,   \quad (j=1,\ldots,n)
\ee
where $n \leq q$, and $U_j=U_j(m^1\ldots,m^n,\lambda) \in \Gamma_0 (V)$ have the following
expansion in the Taylor series around a given $\lambda_0$:
\be \ba{l}   \label{as2}
U_j  = U_j^0 + (\lambda-\lambda_0) \ U_j^1 + (\lambda-\lambda_0)^2 \ U_j^2 + \ldots \ ,  
 \\[2ex]       U^0_j = \e_j B_j \ , \quad U^1_j = \e_j A_j \ , \qquad
A_j \in W  , \quad  B_j \in V  ,  \quad \e_j \in V \ ,
\ea \ee
$A_j, B_j$ are assumed to be invertible, and $\e_1,\ldots,
\e_n$ are mutually orthogonal unit vectors.
We define the discrete net $F$ by the Sym-Tafel formula
\be  \label{as3}
       F = \Psi^{-1} \Psi,_{\lambda} |_{\lambda=\lambda_0} \ ,
\ee
and, finally, we assume that at a single point $m^1_0,\ldots,m^n_0$ (at least) 
we have: 
\[ \Psi (m^1_0,\ldots,m^n_0, \lambda_0) 
\in \Gamma_0  (V) \ ,  \quad  \Psi (m^1_0,\ldots,m^n_0,\lambda) \in  \Gamma_0  (V\oplus W)
\ . \]
Then
\begin{itemize}
\item $(F (m^1,\ldots,m^n) - F (m^1_0,\ldots,m^n_0))  \in V \wedge W$  
\item $F$ is a circular net if and only if for $ k\neq j$
\be      \label{condcir}
A_k (T_k A_j)^{-1} (T_j A_k) A_j^{-1}      \in \R  \ . 
\ee
\end{itemize}
Therefore, $F$ can be always identified with a net in $V \wedge W$.
\end{prop} 

\begin{Proof} 
The compatibility conditions ($T_k T_j \Psi = T_j T_k \Psi$) for the 
linear system \rf{as1}  read
\be \ba{l}   \label{dcc}
(T_k U_j) U_k = (T_j U_k) U_j \ ,
\ea \ee
or, expanding \rf{dcc} in the Taylor series around $\lambda=\lambda_0$,  
\be   \label{cc-dis}  \ba{l}  
 (T_k U^0_j ) U^0_k = (T_j U^0_k ) U^0_j  \ ,  \\[2ex]
(T_k U^1_j ) U^0_k + (T_k U^0_j ) U^1_k  = 
(T_j U^1_k ) U^0_j + (T_j U^0_k ) U^1_j  \ ,  \\[2ex]
(T_k U^2_j ) U^0_k + (T_k U^1_j ) U^1_k  + (T_k U^0_j ) U^2_k = 
(T_j U^2_k ) U^0_j + (T_j U^1_k ) U^1_j  + (T_j U^0_k ) U^2_j \ ,
\ea \ee
and so on. We denote $\Psi_0 := \Psi (m^1,\ldots,m^n,\lambda_0)$.  
Taking into account $U_k (\lambda_0) = U^0_k$, $U_k,_\lambda (\lambda_0) = U^1_k$ and $T_j \Psi_0 = U^0_j \Psi_0$,
we have
\[  \ba{l} \label{DiF}
 D_k F = (T_k \Psi)^{-1} (T_k \Psi),_\lambda |_{\lambda=\lambda_0} - \Psi^{-1} \Psi,_\lambda |_{\lambda=\lambda_0}
= \Psi_0^{-1} (U^0_k)^{-1} U^1_k \Psi_0  \ , \\[2ex]
 T_j D_k F = \Psi_0^{-1} (U^0_j)^{-1} (T_j U^0_k)^{-1} T_j U^1_k  U^0_j  \Psi_0 \ ,  \\[2ex]
(T_k D_j F)^{-1} = \Psi_0^{-1} (U^0_k)^{-1} (T_k U^1_j)^{-1} (T_k U^0_j) U^0_k \Psi_0 \ ,
\ea \]
Applying the first equation of the system \rf{cc-dis} we compute
\[
(T_k D_j F)^{-1} (T_j D_k F) = \Psi_0^{-1} (U^0_k)^{-1} (T_k U^1_j)^{-1}  (T_j U^1_k)  U^0_j  \Psi_0 \ , 
\]
and, finally,
\be
 Q_{kj} (F) =  \\
 \Psi_0^{-1} \left( 
(U^0_k)^{-1} U^1_k (U^0_k)^{-1} (T_k U^1_j)^{-1} (T_j U^1_k)
U^0_j (U^1_j)^{-1} U^0_j \right) \Psi_0 \ ,
\ee
To make further simplification we use \rf{as2} and take into
account that any element of $V$ anti-commutes with any any element of $W$:
\[  
Q_{kj} (F) = - \Psi_0^{-1} B_k^{-2} A_k (T_k A_j)^{-1} (T_j A_k) A_j^{-1} B_j^2 \Psi_0 \ .
\]
Therefore, using $V \perp W$ 
and $\Psi_0 \in \Gamma_0 (V)$ (because $U^0_j \in \Gamma_0  (V)$), 
we get  
 \[
Q_{kj} (F) = - B_k^{-2} B_j^2 \ A_k (T_k A_j)^{-1} (T_j A_k) A_j^{-1} \ .
\]
which ends the proof of  the second statement of the Proposition~\ref{long}.  
To prove the first statement we will show that 
$(T_j F -  F) \in V \wedge W$. Indeed,  
\[
T_j F - F = ((U_j \Psi)^{-1} (U_j \Psi),_{\lambda} - 
\Psi^{-1} \Psi,_\lambda )|_{\lambda=\lambda_0} = 
\Psi_0^{-1} (U^0_j)^{-1} U^1_j \Psi_0 = \Psi_0^{-1} B_j^{-1} A_j  \Psi_0 \ .
\]
To complete the proof we notice that $A_j$ commutes with
any element of $\Gamma_0 (V)$, and $\Psi_0^{-1} B_j^{-1} \Psi_0 \in V$.
Therefore $\Psi_0^{-1} B_j^{-1} A_j  \Psi_0 \in V \wedge W$.
\end{Proof}

\begin{prop}
If $U^2_k = 0$ (in particular, if \  $U_k$ are linear in $\lambda$), then $F$ defined by
\rf{as1}, \rf{as2}, \rf{as3} is a circular net.
\end{prop}
\begin{Proof} We are going to show that in this case the condition \rf{condcir}
follows from the compatibility conditions. Indeed, because  $U^2_k = 0$, then from
\rf{cc-dis} we get $(T_k U^1_j) U^1_k = (T_j U^1_k) U^1_j$, which can be rewritten as
$ (T_k A_j) A_k  = - (T_j A_k)  A_j $. Hence
$  A_k^{-1} (T_k A_j)^{-1} (T_j A_k) A_j = - 1$ and 
$  A_k (T_k A_j)^{-1} (T_j A_k) A_j^{-1} = - A_k^2 A_j^{-2} \in \R$. 
\end{Proof}

\begin{prop}
If $\dim W = 1$, then $F$ defined by
\rf{as1}, \rf{as2}, \rf{as3} is a circular net in $V$.
\end{prop}
\begin{Proof}
If $\dim W = 1$, then the condition \rf{condcir} is obvious, and  $W \wedge V \simeq V$.
\end{Proof}

\begin{prop} If there exists a discrete  net  $F_A : {\mathbf Z}^n \rightarrow \R^m$ such that 
$D_k F_A = A_k$, then Proposition~\ref{long} can be reformulated as follows: 
$F$ is a circular net if and only if $F_A$ is a circular net.
\end{prop}

The reduction to the group ${\rm Spin} (V\oplus W)$ is always possible as is shown
by the following proposition. 

\begin{prop}  \label{spin} 
If $B_j$ and $A_j$ are unit vectors (for $j=1,\ldots,n$) then 
$\Psi_0 \equiv \Psi (m^1,\ldots,m^n,\lambda_0) \in {\rm Spin} (V)$
and $F$ takes values in ${\rm Spin} (V\oplus W)$.
\end{prop}

Actually, simple bivectors of the form $\hat{v} \wedge \hat{w}$ (where $\hat{v}$,  
$\hat{w}$ are unit vectors from $V$) belongs both  to ${\rm Spin} (V)$ and 
to the Lie algebra of ${\rm Spin} (V)$.  Therefore,  $F$ takes values also in the 
Lie algebra of ${\rm Spin} (V\oplus W)$ provided that the assumptions of Proposition~\ref{spin} 
are satisfied.

We recall the so called main anti-automorphism 
$\beta$ of the Clifford algebra (known also as the reversion) \cite{Cliff,Lou}:
\[  \ba{l}
  \beta ({\mathbf v}_1 {\mathbf v}_2 \ldots {\mathbf v}_k ) := 
{\mathbf v}_k {\mathbf v}_{k-1} \ldots {\mathbf v}_1  \ , \\[2ex]
\beta (c_1 X + c_2 Y) = c_1 \beta (X) + c_2 \beta (Y) \ ,
\ea \]
for any  ${\mathbf v}_j \in V$, any $X, Y \in Cl (V)$, and  $c_1, c_2 \in \R$.
 The group ${\rm Spin (V)}$ consists of products of unit vectors which means that
$X \in {\rm Spin (V)}$ if and only if $\beta (X) X = \pm 1$.

The $\Gamma (V\oplus W)$-valued spectral problem given by \rf{as1}, \rf{as2}  
can always  be transformed into ${\rm Spin} (V\oplus W)$-valued spectral problem
$ T_k \Phi = {\hat U}_k \Phi$ 
by the transformation  $\Phi := g \Psi$, where $g:=| \beta (\Psi) \Psi |^{-1/2}$ is 
a real function, and  ${\hat U}_k := g^{-1} (T_k g) U_k$.

Let $P : W \rightarrow \R$ is  a projection (linear homomorphism of 
vector spaces satisfying $P^2 = P$). We extend its action on $V \wedge W$
in a natural way. Namely, if \ ${\mathbf v}_k \in V$ and ${\mathbf w}_k \in W$, then 
\[   \displaystyle
P (\sum_{k} {\mathbf v}_k {\mathbf w}_k) := \sum_{k} P({\mathbf w}_k) {\mathbf v}_k \ .
\]
\begin{prop}  \label{proj}
Let $P$ is a projection and $F$ is defined by \rf{as1}, \rf{as2}, \rf{as3}. Then 
$P (F)$ is a circular net.  
\end{prop}

\begin{Proof} We denote $P(A_k) = a_k \in \R$. To compute $Q_{kj} ( P(F))$ we need:
\[  \ba{l}
 D_k P (F) = P (D_k F) = a_k \Psi_0 B_k^{-1} \Psi_0 \ ,  \\[2ex]
(T_k D_j P (F))^{-1} = (T_k a_j)^{-1} \Psi_0^{-1}  B_k^{-1} \e_k^{-1} (T_k B_j) \e_k B_k \Psi_0 \ ,
\\[2ex]  
T_j (D_k (F)) = (T_j a_k) \Psi_0^{-1} B_j^{-1} \e_j^{-1} (T_j B_k)^{-1} \e_j B_j \Psi_0 \ , \\[2ex]
(D_j P (F))^{-1} = a_j^{-1} \Psi_0^{-1}  B_j \Psi_0 \ .
\ea \]
The compatibility conditions $(T_k U^0_j) U^0_k = (T_j U^0_k) U^0_j$, after taking into 
account $\e_j \e_k^{-1} = - \e_k^{-1} \e_j$, are equivalent to 
$\e_k^{-1} (T_k B_j) \e_k B_k = - \e_j^{-1} (T_j B_k) \e_j B_j$. Therefore
\[
Q_{kj} ( P(F)) = - \frac{ a_k (T_j a_k)}{a_j (T_k a_j)} \Psi_0^{-1} B_k^{-1} B_k^{-1} 
B_j B_j \Psi_0 = - \frac{ a_k (T_j a_k) B_j^2}{a_j (T_k a_j) B_k^2} \in \R \ ,
\] 
which completes the proof (compare \cite{Ci-cross}).  
\end{Proof}

We proceed to several examples, where 
$U_j$ are rational with respect to $\lambda$ (usually 
even linear in $\lambda$). All assumptions of Proposition~\ref{long} are 
assumed to be satisfied. We denote by $\e_1,\ldots,\e_q$ and $\e_{q+1},\ldots,\e_{q+r}$ 
orthonormal bases in $V$ and $W$, respectively.

The corresponding smooth (continuum) cases were considered in \cite{Ci-Mess,Ci-Cliff,Ci-Spin},
and also (by  different approaches) in \cite{ABT,BP-izot,BDPT,FP,HJ-Gui,Sch} 
and earlier \cite{Eis1,Eis2}.

If $U_j$ are linear in $\lambda$, then it is not difficult  
to construct the Darboux-B\"acklund transformation (similarly as in the smooth case, \cite{BC-pla}).
The details, analogical to the special case of discrete isothermic surfaces \cite{Ci-Kent}, 
will be presented elsewhere.

\subsubsection*{Discrete isothermic surfaces in $\R^q$} 

\[  \ba{l}
V \simeq \R^q \ , \quad  W \simeq \R^{1,1} \ , \quad n=2 \ , \quad r=2 \ , \quad  U_j = U^0_j + \lambda U^1_j \ ,  \\[1ex]
P (\e_{q+1}) = 1 \ , \quad P (\e_{q+2}) = \pm 1 \ .
\ea \] 
Smooth isothermic immersions admit isothermic (isometric) parameterization of curvature lines.
In these coordinates $ds^2 = \Lambda ( (dx^1)^2 + (dx^2)^2 )$ and the second fundamental form
is diagonal. 

\subsubsection*{ Discrete Guichard nets in ${\mathbf R}^q$}

\[  \ba{l}
V \simeq \R^q \ , \quad  W \simeq \R^{2,1} \ , \quad n=3 \ , \quad r=3 \ , \quad  U_j = U^0_j + \lambda U^1_j \ ,  \\[1ex]
P (\e_{q+1}) = \cos \varphi_0 \ , \quad P (\e_{q+2}) = \sin \varphi_0 \ , \quad P (\e_{q+3}) = \pm 1 \ , \quad  \varphi_0 = \const \ .
\ea \] 
Guichard nets in $\R^3$ are characterized by 
the constraint  $H_1^2 + H_2^2 = H_3^2$,  where $H_j$ are Lam\'e coefficients, i.e., 
$ds^2 = H_1^2 (dx^1)^2 + H_2^2 (dx^2)^2 + H_3^2 (dx^3)^2$.

\subsubsection*{ Discretization of some class of orthogonal nets in ${\mathbf R}^n$}  
\[  \ba{l}
V \simeq \R^n \ , \quad  W \simeq \R^n \ , \quad q=n \ , \quad r=n \ , \quad  U_j = U^0_j + \lambda U^1_j \ ,  \\[1ex]
P (\e_{n+k}) = 1 \ , \quad P (\e_{n+j}) = 0 \quad (j \neq k) \ .
\ea \] 

This class in the smooth case is defined by the constraint  
$H_1^2 + \ldots + H_n^2 = \const$, where $H_j$ are Lam\'e coefficients, i.e., 
$ds^2 = H_1^2 (dx^1)^2 + \ldots + H_n^2 (dx^n)^2$.

\subsubsection*{Discrete Lobachevsky $n$-spaces in $\R^{2n-1}$}
\[  \ba{l}
V = V_1 \oplus V_2 \ , \quad V_1 \simeq \R^n \ , \quad V_2 \simeq \R^{n-1} \ , \quad  W \simeq \R \ , \quad \lambda_0 = 1 \ , \\[2ex]
 U_j = \e_j \left( \frac{1}{2} \left( \lambda - \frac{1}{\lambda} \right) A_j + 
\frac{1}{2} \left( \lambda + \frac{1}{\lambda} \right) P_j + Q_j \right) \ ,  \\[2ex]
\e_j \in V_1 \ , \quad Q_j \in V_1 \ , \quad P_j \in V_2 \ , \quad A_j \in W \ ,  \quad
P_j + Q_j = B_j \ .
\ea \] 
In the continuum limit we get immersions with the constant negative sectional curvature 
(Lobachevsky spaces) \cite{Am77,Am80,TT}. The discrete case is presented in more detail 
in \cite{Ci-space}.

In all presented cases the continuum limit, done by assuming 
$\varepsilon \rightarrow 0$ where $\varepsilon$ is the size of the 
${\bf Z}^n$ lattice (compare \cite{BMS,BP-izot}), 
seems to be correct
because all algebraic properties (including the integrability) 
are preserved by our  
discretization. 

However, a purely geometrical characterization is known only in the case 
of isothermic surfaces: the cross ratio is harmonic \cite{BP-izot}.
Therefore  a  geometrical characterization 
of discrete nets presented in my paper is 
an important open problem. The general characterization 
of families of 
smooth submanifolds corresponding to the discrete nets described in Proposition~\ref{long}
(see also \cite{Ci-Spin}) is also not done. 

{\it Acknowledgements}. I was prompted to complete this work by Prof.\ I.\ Kh.\ Sabitov 
during the Research Group 
``Differential equations in the geometry of submanifolds and mathematical physics''
(organized 8-19.11.2004 by Prof.\ Yu.\ A.\ Aminov at the Banach Center in Warsaw, Poland)
where I presented the results of \cite{Ci-Spin}.

\end{document}